\documentclass[lnicst]{svmultln}
\usepackage{makeidx}  
\usepackage{amsfonts}
\usepackage{amsmath,cases,amssymb}
\usepackage[dvips]{graphicx}
\usepackage{cite,color}
\usepackage{color}
\usepackage{epsfig}
\usepackage{epstopdf}
\usepackage{url}
\usepackage{algorithm}
\usepackage{algorithmicx, algpseudocode}
\usepackage{color}
\usepackage{multirow}
\usepackage{mathtools}
\usepackage{balance}
\usepackage{appendix}
\newcommand{\rank}{{\sf rank}}

\newcommand{\clL}{{\cal L}}

\newcommand{\ds}{\displaystyle}

\newcommand{\clF}{{\cal F}}

\newcommand{\clN}{{\cal N}}
\newcommand{\clG}{{\cal G}}

\newcommand{\Tr}{\mbox{Trace}}

\newtheorem{mypro}{Proposition}
\newtheorem{mylem}{Lemma}

\newcommand{\bgeqn}{\begin{equation}}
\newcommand{\edeqn}{\end{equation}}
\newcommand{\la}{\langle}
\newcommand{\ra}{\rangle}

\newcommand{\beqa}{\begin{eqnarray}}
\newcommand{\eeqa}{\end{eqnarray}}
\newcommand{\beqas}{\begin{eqnarray*}}
\newcommand{\eeqas}{\end{eqnarray*}}

\newcommand{\clT}{{\cal T}}

\newcommand{\clV}{{\cal V}}
\newcommand{\clW}{{\cal W}}
\newcommand{\clR}{{\cal R}}

\begin{document}
\mainmatter              
\title{Mixed integer nonlinear programming for Joint Coordination of Plug-in Electrical Vehicles Charging and Smart Grid Operations}
\titlerunning{Mixed integer nonlinear programming}  
%
\author{Y. Shi\inst{1} \and H. D. Tuan\inst{1}
\and A. V. Savkin\inst{2}}
\authorrunning{Y. Shi, H. D. Tuan, A. V. Savkin}   
\institute{
University of Technology Sydney, Broadway, NSW 2007, Australia\\
\email{Ye.Shi@student.uts.edu.au, Tuan.Hoang@uts.edu.au},
\and The University of New South Wales, Sydney, NSW 2052, Australia\\
\email{a.savkin@unsw.edu.au}}

\maketitle              

\begin{abstract}        
The problem of joint coordination of plug-in electric vehicles (PEVs) charging and grid power control is to minimize both PEVs charging cost and energy generation cost, while meeting both residential
 and PEVs' power demands and suppressing the potential impact of  PEVs integration.
A bang-bang PEV charging strategy is adopted to exploit its simple online implementation, which
 requires computation of  a mixed integer nonlinear programming problem (MINP) in binary variables of the PEV charging strategy and continuous variables of the grid voltages. A new  solver for this MINP is proposed.
 Its efficiency is shown by numerical simulations.
\end{abstract}

\section{Introduction}

In recent years, there has been increasing concern over the energy consumption and environment pollution. On the other hand, advance in the battery and smart grid technology have drawn growing attention to electric vehicles (EVs)\cite{EEI2014}. It is expected that more than 50\% of the new vehicles will be EVs by 2020 \cite{SGT_book}. The massive penetration of plug-in electric vehicles (PEVs) can pose potential threats to the stability of a power grid, which is not easily compensated \cite{Laetal16}. Unregulated charging of PEVs
may cause overloading, additional power loss and unacceptable voltage violation \cite{HSX12}. Coordinate PEV charging
is thus needed for the cost-saving services for PEVs and meeting PEV power demands and operation constraints in smart grid system.

To address the coordinate PEV charging problem, \cite{FRR15} proposed a mixed integer nonlinear programming (MINLP) model in an unbalanced system. This MINLP model was then linearized to mixed integer linear programming (MILP) model through the first order Taylor expansion and piecewise linear approximation. As a result, the solution of MILP is not necessarily feasible to the original MINLP problem. Adding a vehicle-to-grid (V2G) charging strategy, similar MILP model was proposed in \cite{AFRR16}. Nevertheless, the practicability of PEVs' discharging involving costs and technologies raised continuous concern in \cite{CHD11}. In addition, the above references for coordinate PEV charging are all based on the off-line strategy. In that case, prior information including the arrival time, departure time and state of charge (SoC) of PEVs must be given beforehand. It is not practical to obtain all of those information in advance. To deal with the online PEV coordination, model predictive control (MPC) approach has been widely used in recent studies\cite{RSME16,TZ17}. A MPC-based model proposed in \cite{TZ17} scheduled PEV charging in a finite horizon, but operation constraints of grid were not considered. A MILP model over a rolling horizon window for energy storage control was
developed in \cite{MSE14} with power balance constraints ignored. Reference \cite{RSME16} proposed a stochastic optimization algorithm to tackle the MILP-based MPC model for different types of PEVs coordination problems, which suffers from large computational cost.

In this paper a bang-bang strategy is adopted
for PEV charging. At each time slot individual PEVs either charge at a maximal power rate
or do not charge at all. The obvious merit of such bang-bang charging strategy is its easy and efficient online implementation. At each time slot, it requires a joint PEV charging coordination and grid power control for a model predictive system, which is a MINLP in the bang-bang PEV charging variables and the bus voltage variables of the grid.
A new approach is developed to handle this MINLP. Firstly, by relaxing the nonlinear constraints of the node voltage variables, the MINLP is convexified to a mixed integer convex programming (MICP). Then we develop a new path-following algorithm for computation of this MICP. The found binary value of the PEV charging
coordination is then substituted to the original MINLP for optimizing the bus voltage variables,
for which our previously developed nonconvex spectral optimization algorithm \cite{Phetal12,Yeetal17} is ready for solution. Simulations show that the proposed approach is capable of locating
the optimal solution of this MINLP.

The rest of the paper is structured as follows. Section
II is devoted to an MINLP-based model for the joint coordination of  bang-bang PEV charging and grid power control with analysis on its computational challenges. Section III develops a solver for this MINLP. Simulations are provided in Section IV. Section V concludes the paper.

\section{MPC for joint PEV bang-bang charging coordination and grid power control}
Like \cite{Yeetal17s}, we consider an electric power grid with a set of buses ${\cal N} := \{1, 2,..., N\}$ connected
through a set of flow lines ${\cal L}\subseteq {\cal N}\times {\cal N}$, under which
bus $k$ is connected to bus $m$ if and only if $(k,m)\in {\cal L}$. Denote by  $\clN(k)$
the set of other buses connected to bus $k$. ${\cal G}\subseteq {\cal N}$
is the set of those buses that are connected to distributed generators (DGs).
Bus $k\in {\cal N}\setminus {\cal G}$ is not connected to DGs and bus $k\in {\cal G}$ also has a function
to serve PEVs and will be  referred as charging station (CS) $k$. Thus, there are $M=|{\cal G}|$  CSs in the grid.
The serving time period of the grid is divided into $T$ time slots $\clT:=\{1, 2,\dots, T\}$.

Denote by ${\cal H}_k$ the set of those PEVs that arrive at CS $k$. Accordingly, $k_n$ is the $n$-th PEV that arrives at CS $k$. Each PEV $k_n$ arrives at $t_{a,k_n}\in \clT$ and requires to be fully
charged by its departing time $t_{k_n,d}\in\clT$.
Suppose that $C_{k_n}$ and $s_{k_n}^0$ are the battery capacity and initial SOC of PEV $k_n$ while
$\bar{P}_{k_{n}}$ is the maximum power that its battery can charge during one time slot.
In this paper, we adopt the bang-bang charging strategy, under which  PEV $k_n$
either charges the maximal power  $\bar{P}_{k_{n}}$ or does not charge at all at each time slot.
We use the binary variable
\begin{equation}\label{binary}
\tau_{k_n}(t')\in\{0, 1\}
\end{equation} to implement this strategy, i.e.
PEV $k_n$ charges the power  $P_{k_n}(t')=\tau_{k_n}(t')\bar{P}_{k_{n}}$ during the time slot $t'$.
Accordingly, the following constraint enables PEV $k_n$ to be fully charged at its departure:
\begin{eqnarray}\label{ebi2}
\ds\sum_{t'=t_{k_n,a}}^{t_{k_n,d}}\tau_{k_n}(t')=\bar{\tau}_{k_n},
\end{eqnarray}
where $u_h$ is the charging efficiency of the battery, $\bar{\tau}_{k_n}:=\lceil \frac{ C_{k_n}(1 - s_{k_n}^{0})}{ u_h\bar{P}_{k_n}}\rceil$. For ease of presentation, we set $\tau_{k_n}(t')=0$ for  $t'\notin [t_{k_n,a}, t_{k_n,b}]$.

From the grid side, let $y_{km}\in\mathbb{C}$ be the admittance of line $(k,m)$, The current $I_k(t')$ at node $k\in \clN$ is
$V_k(t')$ is the complex voltage at bus $k$ during the time slot $t'$,
the total supply and demand energy is balanced as:
\begin{eqnarray}
 V_k(t')[\sum_{m\in \clN(k)} y_{km} (V_k - V_m)]^* =  [P_{g_{k}}(t') - P_{l_{k}}(t')
  -\ds\sum_{n\in {\cal H}_k}\bar{P}_{k_n}\tau_{k_n}(t')] \nonumber \\
  + j[Q_{g_{k}}(t') - Q_{l_{k}}(t')], k\in \clG, \label{PEV1h}\\
V_k(t)\ds[\sum_{m\in \clN(k)} y_{km} (V_k - V_m)]^*=- P_{l_{k}}(t')  - j Q_{l_{k}}(t'),\quad  k\in {\cal N}\setminus {\cal G}, \label{PEV1i}
\end{eqnarray}
where $P_{l_{k}}(t')$ and $Q_{l_{k}}(t')$ are respectively known real  and reactive price-inelastic demands to express
the residential power demand, $P_{g_k}(t')$ and $Q_{g_k}(t')$ are the real and reactive powers generated by DG $k$.

The following standard constraints are also set.
\begin{itemize}
\item The range of generated powers by the DGs:
\begin{eqnarray}
{\underline P}_{g_k} \leq P_{g_k}(t') \leq {\overline P}_{g_k},\quad {\underline Q}_{g_k} \leq Q_{g_k}(t') \leq {\overline Q}_{g_k}, \quad k\in {\cal G}, \label{PEV1d}
\end{eqnarray}
where ${\underline P}_{g_k}$, ${\underline Q}_{g_k}$ and ${\overline P}_{g_k}$, ${\overline Q}_{g_k}$ are respectively
lower and upper physical limits of the real generated  and reactive generated powers.
\item Voltage range and phase balance:
\begin{eqnarray}
{\underline V}_k \leq |V_k(t')| \leq {\overline V}_k,\quad
|\mbox{arg}(V_k(t'))-\mbox{arg}(V_m(t'))| \leq \theta_{km}^{\max}, \label{PEV1g}\\
 k\in {\cal N}, (k,m)\in\clL, t'\in\clT, \nonumber
\end{eqnarray}
where ${\underline V}_k$ and ${\overline V}_k$ are the lower limit and upper limit of the voltage amplitude,
while $\theta^{\max}_{k,m}$ are given to express the voltage phase balance.

\end{itemize}

The cost function is defined as the sum of the energy cost to DGs and charging cost for PEVs
\begin{equation}\label{objective1}
\clF(\clR,\tau) = \ds \sum_{t'\in \clT} \sum_{k\in {\cal G}}f(P_{{g_k}}(t')) + \ds \sum_{t'\in \clT} \sum_{k\in {\cal N}} \sum_{n\in {\cal H}_k} \beta_t\tau_{k_n}(t')\bar{P}_{{k_n}},
\end{equation}
where $f(P_{{g_k}}(t'))$ is the cost function of real power generation  by DGs, which is linear or quadratic in $P_{{g_k}}(t')$,
and $\beta_t$ is the known PEV charging price during the time slot $t'$.

By defining
$R(t')=\{P_g(t'), Q_g(t')\}, \clR=\{R(t')\}_{t'\in\clT}$,
and $\tau=\{\tau(t')\}_{t'\in\clT},\\
\tau(t')=\{\tau_{k_n}(t')\}_{k_n\in{\cal H}_k}$,
$(R(t'), \tau^{PEV}(t'))$ and $V(t')$ are considered as the system state and control, respectively.
As such, the joint PEV charging coordination and voltage control to optimize the energy and charging costs appears to be the following control problem over the finite horizon $[1,T]$:
\begin{equation}\label{PEV1}
\ds\min_{\clV, \clR,\tau^{PEV}}\ \clF(\clR,\tau^{PEV})\quad
\mbox{s.t.}\quad  (\ref{binary}), (\ref{ebi2}), (\ref{PEV1h})-(\ref{PEV1g}).
\end{equation}
However, all equations in (\ref{PEV1}) are not known a priori.

Denote by $C(t)$ the set of PEVs that need to be charged at $t$ and ahead. For each $k_n\in C(t)$, let
$d_{k_n}(t)$ be its remaining demand for charging by the departure time $t_{k_n,d}$. Therefore,
the binary variable
\begin{equation}\label{binarya}
\tau_{k_n}(t')\in \{0,1\}, t'\in [t, k_{k_n,d}], k_n\in C(t)
\end{equation} must satisfy the following constraints:
\begin{equation}\label{binary1}
\sum_{t'=t}^{t_{k_n,b}}\tau_{k_n}(t')=\bar{\tau}_{k_n}(t),\ k_n\in C(t),
\end{equation}
where $\bar{\tau}_{k_n}(t):={\LARGE \lceil}\frac{d_{k_n}(t)}{u_h\bar{P}_{k_n}}{\LARGE \rceil}$.
Define $
\Psi(t)=\max_{k_n\in C(t)} t_{k_n,d}$,
we propose an online algorithm, which at time $t$ solves the following MPC over the prediction horizon
$[t,\Psi(t)]$ but then takes only
$V(t), R(t)$ and $\tau(t)$, for updating the solution of (\ref{PEV1}):
\allowdisplaybreaks[4]
\begin{subequations}\label{HOR1}
\begin{eqnarray}
\ds\min_{\clV_P(t), \clR_P(t), \tau_P(t)} F_P(\clR_P(t),\tau_P(t))\quad \nonumber
\mbox{s.t.}\quad (\ref{PEV1i})-(\ref{PEV1g}), (\ref{binarya}), (\ref{binary1}), \label{HOR1a}\\
 V_k(t')\ds[\sum_{m\in \clN(k)} y_{km} (V_k(t') - V_m(t'))]^* =
[P_{g_{k}}(t') - P_{l_{k}}(t') \nonumber \\
 -\sum_{k_n\in C(t)}\bar{P}_{k_n}\tau_{k_n}(t')] + j(Q_{g_{k}}(t') - Q_{l_{k}}(t')),\quad (t',k)\in [t,\Psi(t)]\times \clG. \label{HOR1b}
\end{eqnarray}
\end{subequations}
One can see (\ref{HOR1}) is  a difficult MINP  because (\ref{PEV1i}), (\ref{PEV1g}) and (\ref{HOR1b})
are nonlinear in the voltage variable $V(t')$ while (\ref{binarya}) is a discrete combinatoric constraint.
In the next section, we propose an efficient  approach, which also exploits the fact that only the snapshot at $t$ of
the solution of (\ref{HOR1}) is extracted to update the online solution of (\ref{PEV1}).
\section{Solver for MINP}
For  $W(t') := V(t')V^H(t') \in \mathbb{C}^{N\times N}$,
which must satisfy $ W(t')\succeq 0$ and $\rank(W(t'))=1$,
we replace  $W_{km}(t')=V_k(t')V^*_m(t')$,  $(k,m)\in\clN\times\clN$ in,
in (\ref{HOR1}) to  reformulate it to the following MINP  in  matrix-valued variable $\clW_P(t):=\{W(t')\}_{t'\in [t,\Psi(t)]}$ and binary-valued variable $\tau_P(t)$:
\allowdisplaybreaks[4]
\begin{subequations}\label{rHOR1}
\begin{eqnarray}
\ds\min_{\clW_P(t), \clR_P(t), \tau_P(t)} F_P(\clR_P(t),\tau_P(t)) \quad \nonumber
\mbox{s.t.}\ (\ref{PEV1d}), (\ref{binarya}), (\ref{binary1}), \label{rHOR1a}\\
\ds\sum_{m\in \clN(k)} (W_{kk}(t') - W_{km}(t'))y_{km}^*  =
 [P_{g_{k}}(t') - P_{l_{k}}(t') \nonumber \\
 -\sum_{k_n\in C(t)}\bar{P}_{k_n}\tau_{k_n}(t')]
 + j(Q_{g_{k}}(t') - Q_{l_{k}}(t')),  \quad  k\in \clG,  \label{rHOR1b}  \\
\sum_{m\in \clN(k)}(W_{kk}(t') - W_{km}(t'))y_{km}^* =
 - P_{l_{k}}(t')  - j Q_{l_{k}}(t'), k\notin  {\cal G}, \label{rHOR1c}\\
{\underline V}_k^2 \leq W_{kk}(t') \leq {\overline V}_k^2, \quad  k\in {\cal N}, \label{rHOR1f}\\
\Im(W_{km}(t'))\leq \Re(W_{km}(t'))\tan(\theta_{km}^{max}), (k,m)\in\clL,  \label{rHOR1g}\\
W(t')\succeq 0,  \label{rHOR1k}\\
\mbox{rank}(W(t'))=1, \label{rHOR1l}
\end{eqnarray}
\end{subequations}
The difficulty of (\ref{rHOR1}) is concentrated on the multiple nonconvex  matrix rank-one constraints in
(\ref{rHOR1l}) and multiple binary  constraints in (\ref{binarya}).
Below we propose a two-stage optimization approach to its online algorithm.
In the first optimization stage, we drop the matrix rank-one constraint (\ref{rHOR1l}) to relax (\ref{rHOR1}) to
the following MICP for $t'\in [t,\Psi(t)]$:
\begin{eqnarray}\label{rHOR2}
\ds\min_{\clW_P(t), \clR_P(t), \tau_P(t)} F_P(\clR_P(t),\tau_P(t))\quad
 \mbox{s.t.}\ (\ref{PEV1d}),(\ref{binarya}), (\ref{binary1}), (\ref{rHOR1b})-(\ref{rHOR1k}).
\end{eqnarray}
Suppose that $(\hat{\clW}_P(t)$, $\hat{\clR}_P(t))$ and $\hat{\tau}_{P}(t)$ is the optimal solution of (\ref{rHOR2}). If
$\mbox{rank}(\hat{W}(t'))\equiv 1$, $t'\in [t,\Psi(t)]$, then $\hat{V}(t')$ such that $\hat{W}(t')=\hat{V}(t')\hat{V}^H(t')$
together with $\hat{R}(t')$ and  $\hat{\tau}_{k_n}(t')$ constitute
the optimal solution of MINP (\ref{HOR1}). Otherwise, we go to the next optimization stage, which
substitutes $\hat{\tau}_{k_n}(t)$ into (\ref{rHOR1b}) to consider the snapshot at $t$ of (\ref{rHOR1}) only
\begin{subequations}\label{rHOR6}
\begin{eqnarray}
\ds\min_{W(t), R(t)}\ F(P_g(t))):=\ds \sum_{t'=t}^{\Psi(t)}
\sum_{k\in {\cal G}}f(P_{{g_k}}(t'))\quad
 \mbox{s.t.}\quad (\ref{PEV1d}), (\ref{rHOR1c})-(\ref{rHOR1k})\label{rHOR6a}\\
\ds\sum_{m\in \clN(k)} (W_{kk}(t) - W_{km}(t))y_{km}^*  =
 [P_{g_{k}}(t) - P_{l_{k}}(t) \nonumber \\
 -\sum_{k_n\in C(t)}\bar{P}_{k_n}\hat{\tau}_{k_n}(t)]
 + j(Q_{g_{k}}(t) - Q_{l_{k}}(t)),  \quad  k\in \clG,  \label{rHOR6b}  \\
 \mbox{rank}(W(t))=1,\label{rHOR6c}
\end{eqnarray}
\end{subequations}
which involves  only one matrix rank-one constraint (\ref{rHOR6c}). The rationale behind this simplified treatment is that
in the end we need only the snapshot at $t$ of the  solution of (\ref{rHOR1}) for online updating the
voltage $V(t)$ and generated power $R(t)$.

In the next two subsections we propose algorithms for solving  MICP  (\ref{rHOR2}) and the nonconvex optimization
problem (\ref{rHOR6}).
\subsection{New computational solution for MICP problem (\ref{rHOR2})}
It is clear that the main task is how to cope with the discrete constraint (\ref{binarya}) in MICP (\ref{rHOR2}). The following result establishes the equivalence between this discrete constraint and a set of continuous constraints.
\begin{mylem}\label{bilem} Under the linear constraint (\ref{binary1}), the binary constraint (\ref{binarya})
is equivalent to the following set of continuous constraints:
\begin{eqnarray}
 0\leq \tau_{k_n}(t')\leq 1, t'\in [t,t_{k_n,d}], k_n\in C(t),\label{aia}\\
 g(\tau_P(t))\geq \bar{\tau}(t):=\sum_{k_n\in C(t)}\bar{\tau}_{k_n}(t), \label{ai}
\end{eqnarray}
where $g(\tau_P(t)):=\sum_{k_n\in C(t)}\sum_{t'=t}{t_{k_n,b}}\tau^L_{k_n}(t')$, for $L>1$.
\end{mylem}
The following result is a direct consequence of Lemma \ref{bilem}.
\begin{mypro}\label{bilem1} Under the linear constraint (\ref{binary1}), the function
\[
g_1(\tau_P(t)):=\frac{1}{g(\tau_P(t))}-\frac{1}{\bar{\tau}(t)}
\]
can be used to measure the degree of satisfaction of
the binary constraint (\ref{binarya}) in the sense that $g_1(\tau_P(t))\geq 0\ \forall\
\tau_{k_n}(t')\in [0,1]$ and $g_1(\tau_P(t))=0$ if
and only if $\tau_{k_n}(t')$  are binary (i.e. satisfying (\ref{binarya})).
\end{mypro}
Therefore MICP (\ref{rHOR2}) is equivalent to the following penalized optimization problem:
\begin{eqnarray}\label{quad1}
\ds\min_{\clW_P(t), \clR_P(t), \tau_P(t)}  \Phi(\clR_P(t),\tau_P(t)):=
F_P(\clR_P(t),\tau_P(t))+
\mu g_1(\tau_P(t)) \nonumber \\
\mbox{s.t.}\quad (\ref{PEV1d})
 \mbox{for}\ t'\in [t,\Psi(t)],  (\ref{binary1}),
 (\ref{rHOR1b})-(\ref{rHOR1k}), (\ref{aia}),
\end{eqnarray}
where $\mu>0$ is a penalty parameter. 
As the function $g(\tau_P(t))$ is convex, it is true that at $\tau^{(\kappa)}_P(t)$ \cite{Tuybook},
\begin{eqnarray}\label{up_bound}
g(\tau_P(t))&\geq&g^{(\kappa)}(\tau_P(t))\nonumber\\
&:=& g(\tau^{(\kappa)}_P(t))+\la \nabla g(\tau^{(\kappa)}_P(t),\tau_P(t)-
\tau^{(\kappa)}_P(t)\ra\nonumber\\
&=&-(L-1)\ds\sum_{k_n\in C(t)}\sum_{t'=t}^{t_{k_n,d}}(\tau^{(\kappa)}_{k_n}(t'))^L+
L\sum_{k_n\in C(t)}\sum_{t'=t}^{t_{k_n,d}}(\tau^{(\kappa)}_{k_n}(t'))^{L-1}\tau_{k_n}(t').\nonumber\\
\end{eqnarray}
Therefore, an upper bounding approximation at $\tau^{(\kappa)}_P(t)$ for $g_1(\tau_P(t))$ can be easily obtained as
\begin{eqnarray}
g_1(\tau_P(t))\leq g_1^{(\kappa)}(\tau_P(t))
:=\ds\frac{1}{g^{(\kappa)}(\tau_P(t))}-\frac{1}{\bar{\tau}(t)}\label{quad2bb}
\end{eqnarray}
over the trust region
\begin{equation}\label{quad2b}
g^{(\kappa)}(\tau_P(t))>0.
\end{equation}
Accordingly, at the $\kappa$-th iteration we solve the following convex optimization problem to generate the
next iterative point $(\clW^{(\kappa+1)}_P(t), \clR^{(\kappa+1)}_P(t), \tau^{(\kappa+1)}_P(t))$:
\begin{eqnarray}\label{quad2}
\ds\min_{\clW_P(t), \clR_P(t), \tau_P(t)}  \Phi^{(\kappa)}(\clR_P(t),\tau_P(t)):= F_P(\clR_P(t),\tau_P(t))
+\mu g_1^{(\kappa)}(\tau_P(t))
\nonumber\\
\mbox{s.t.}\quad (\ref{PEV1d}), \quad \mbox{for}\ t'\in [t,\Psi(t)], (\ref{binary1}),
 (\ref{rHOR1b})-(\ref{rHOR1k}), (\ref{aia}), (\ref{quad2b}).
\end{eqnarray}
We thus arrive at
$\Phi(\clR^{(\kappa+1)}_P(t),\tau_P^{(\kappa+1)}(t))\leq \Phi^{(\kappa)}(\clR^{(\kappa+1)}_P(t),\tau_P^{(\kappa+1)}(t))
<\Phi^{(\kappa)}\\(\clR^{(\kappa)}_P(t),\tau_P^{(\kappa)})
=F(\clR^{(\kappa)}_P(t),\tau_P^{(\kappa)}(t))$,
implying that $\tau_P^{(\kappa+1)}(t)$ is a better feasible point than $\tau_P^{(\kappa)}(t)$ for (\ref{quad1}). For a sufficiently large
$\mu>0$, $g(\tau_P^{(\kappa)}(t))\rightarrow 0$ as well, yielding an optimal solution of  MICP (\ref{rHOR2}).
Pseudo-code for this computational procedure is provided by
Algorithm \ref{alg1}.
\begin{algorithm}
\caption{MICP Solver} \label{alg1}
\begin{algorithmic}
\State{\it Initialization.} Choose a feasible point $\tau_P^{(0)}(t)$
for (\ref{quad1}) as the optimal solution of the following problem by relaxing the binary constraints
(\ref{binarya}) in (\ref{rHOR2}) to box constraints:
\begin{eqnarray}\label{inial1}
\ds\min_{\clW_P(t), \clR_P(t), \tau_P(t)} F_P(\clR_P(t),\tau_P(t))
 \mbox{s.t.}\ (\ref{PEV1d}),  (\ref{rHOR1b})-(\ref{rHOR1k}), (\ref{binary1}),
\tau_{k_n}(t')\in [0,1],
\end{eqnarray}
Set $\kappa=0$. \State{\it $\kappa$-th iteration.}
Solve (\ref{quad2}). If the optimal solution $\tau_P^{(\kappa+1)}(t) $
satisfies $\sum_{k_n\in C(t)}\sum_{t'=t}^{t_{k_n,d}} \left( \tau_{k_n}^{(\kappa+1)}(t') - \left(\tau_{k_n}^{(\kappa+1)}(t') \right)^L \right)\approx 0$,
terminate the algorithm and output $\tau_P^{(\kappa+1)}(t)$ as a found solution.
Otherwise,  reset $\kappa+1 \to \kappa$ and $\tau_P^{(\kappa+1)}(t) \to \tau_P^{(\kappa)}(t)$ for the next
iteration.
\end{algorithmic}
\end{algorithm}

\subsection{Computational procedure for (\ref{rHOR6})}
Following our previous works \cite{Phetal12,STST15,Yeetal17,STA17,Naetal17},
the matrix rank-one constrained optimization problem
(\ref{rHOR6}) is solved via the following penalized optimization problem for $\lambda>0$:
\begin{subequations}\label{r7}
\begin{eqnarray}
\ds\min_{W(t), R(t)}\ F(P_g(t))+ \lambda(\Tr(W(t))-\lambda_{\max}(W(t))),\label{r7a}\\
 \mbox{s.t.}\quad (\ref{PEV1d}), (\ref{rHOR1c})-(\ref{rHOR1k})\quad \mbox{for}\ t'=t,
 \label{r7b}
\end{eqnarray}
\end{subequations}
which is computed by solving the following  convex optimization problem at the $\kappa$th iteration
to generate $W^{(\kappa+1)}(t)$:
\begin{eqnarray}\label{rW5}
\ds\min_{W(t), R(t)} F(P_g(t))+\lambda (\Tr(W(t))\nonumber\\
 -(w_{\max}^{(\kappa)}(t))^H W(t) w_{\max}^{(\kappa)}(t))\quad\mbox{s.t.}\quad (\ref{r7b}),
\end{eqnarray}
where $W^{(k)}(t)$ is a point found from the $(\kappa-1)$th iteration and
$w^{(\kappa)}_{\max}(t)$ denotes the normalized eigenvector
corresponding to the maximal eigenvalue $\lambda_{\max}(W^{(\kappa)}(t))$ of $W^{(\kappa)}(t)$. The rationale behind
using the penalized optimization problem (\ref{r7}) is that $\Tr(W(t))-\lambda_{\max}(W(t))$
is the degree of satisfaction of the matrix rank-one
constraint (\ref{rHOR6c}). The reader is  referred to \cite{Yeetal17} for  proof of its convergence.

\section{Simulation results}

Sedumi\cite{S98} solver under the framework of CVX\cite{cvx}
on a Core i7-7600U processor is used to solve convex optimization problems such as
(\ref{quad2}) and (\ref{rW5}). Simulations are tested on IEEE-30 network, whose structure, physical limits and cost functions $f(P_{{g_k}}(t'))$ are provided in the Matpower library \cite{ZMT11}.
%

The considered charging period is  from 6:00 pm to 6:00 am of the next day to reflect the fact that most PEVs are charged after their owners' working hours. This time period  is  uniformly divided into $24$ 30-minute time slots.
The PEVs  arrive during the time period from 6:00 pm to midnight. The arrival times of PEVs are independent and are generated by a truncated  normal distribution $(8,1.5^2)$.
The battery capacity $C_{k_n}=100$ KWh of PEVs is that of Tesla Model S.
The initial SOC $s^0_{k_n}$ of all PEVs
is set as 20\%.

The price-inelastic load $P_{l_k}(t)$ is defined as
$P_{l_k}(t) = l(t)\times \bar{P}_{l_k}\times T/\sum_{t=1}^{24} l(t)$,
where $\bar{P}_{l_k}$ is the load demand specified by \cite{ZMT11} and $l(t)$ is the residential load demand taken
from the UK \cite{demand_uk}. The time varying energy price is taken from \cite{price_uk} on different days in 2017.
%
The tolerance $\epsilon = 10^{-4}$ is set for the stop criteria and $L=1.5$ in (\ref{up_bound}) is chosen to accelerate the convergence speed for the optimization algorithms.

The computational results are summarized in Table \ref{mpc_network}.
\begin{table*}[h]
    \centering
    \caption{Online bang-bang charging computational results}
    \begin{tabular}{ccccccccc}
    \hline
    \hline
     Networks& Profiles&Binary variables &$\mu$ & $\lambda$ &Obj. (\ref{objective1}) by (\ref{rHOR2})
      &Obj. (\ref{objective1}) by (\ref{rHOR6})& Time (s) \\
    \hline
    \multirow{4}{*}{Case30}& Profile 1 & 3012 &10 &1 & 6560.7&6560.8&  24.0 \\
    & Profile 2 & 3012 &10 &1 & 6569.6&6569.7&  23.1 \\
    & Profile 3 & 3012 &10 &1 & 6615.1&6615.2&  25.2 \\
    & Profile 4 & 3012 &10 &1 & 6589.3&6599.1&  29.4 \\
    \hline
    \hline
    \end{tabular}
\label{mpc_network}
\end{table*}
Its third column provides the number of binary variables $\tau_{k_n}(t')$ in (\ref{PEV1}). The values of
the penalty parameters $\mu$ in  (\ref{quad2}) and $\lambda$ in (\ref{rW5}) are specified
in the forth  and fifth columns. The value of the cost objective (\ref{objective1})
by (\ref{rHOR2})  and (\ref{rHOR6}) are given by the sixth
and seventh columns, respectively. The effectiveness of using (\ref{rHOR6}) is confirmed by observing that these
values are almost the same.
The average running time of computation at each time slot  is shown in the last column.

\begin{figure}[h]
\centering
\includegraphics[width=0.6 \columnwidth]{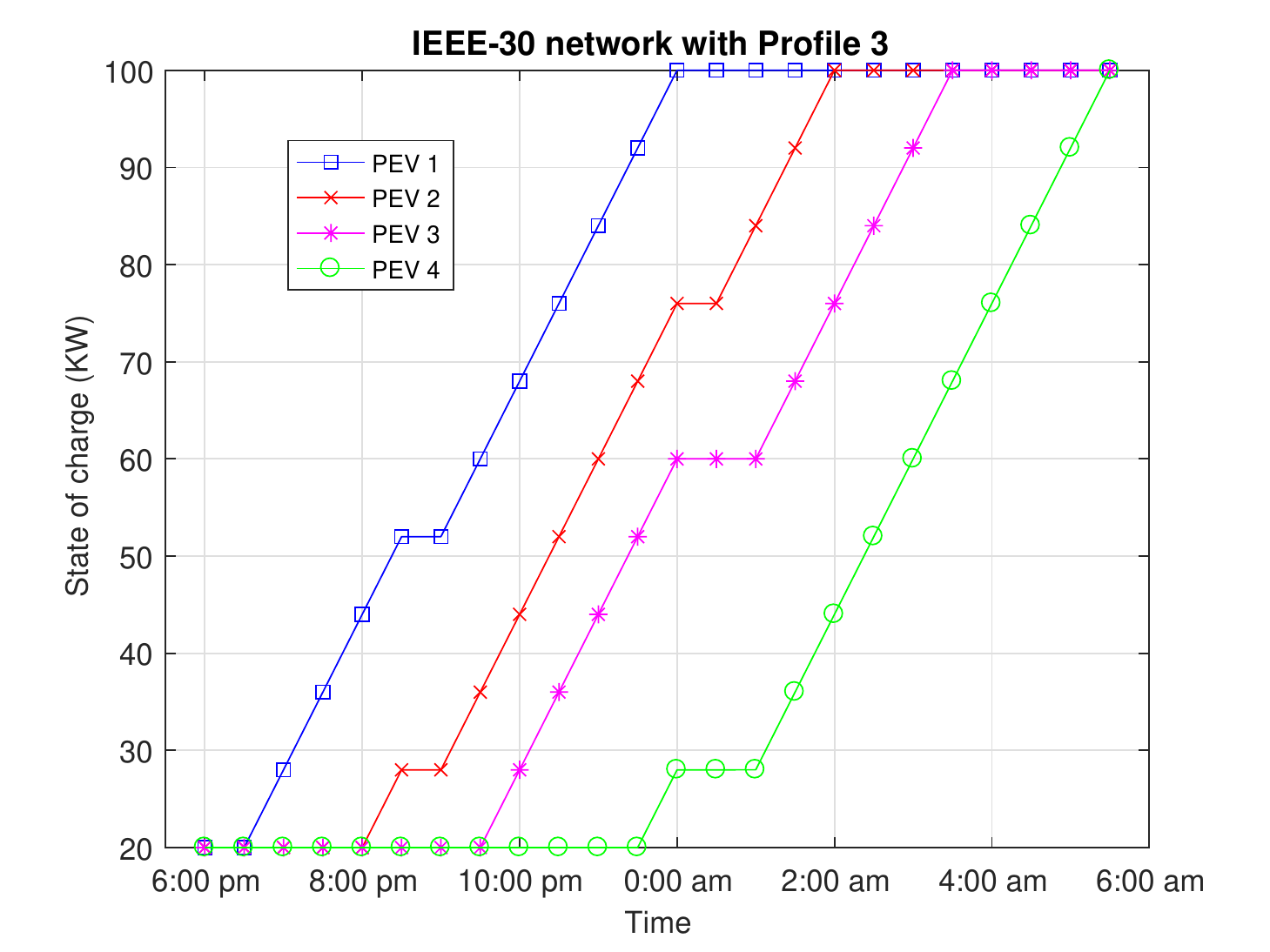}
\caption{The SOC of PEVs during the charging period}
\label{charging_pattern}
\end{figure}

Fig.\ref{charging_pattern} plot the SoC of  four PEVs randomly taken
from simulation on profile 3, which arrive at different times. For a few time slots, PEVs do not charge so their SoC remain unchanged.

\section{Conclusions}
The joint online coordination of PEV bang-bang charging  and power control
to serve both PEVs at a competitive cost and residential power demands at a competitive operating cost
is very difficult due to the random nature of PEVs' arrivals and  demands and the discrete nature of bang-bang charging.
We have proposed a novel and easily-implemented  MPC-based two-optimization stage online algorithm that can achieve an optimal solution.

%

\end{document}